\documentclass[preprint2]{aastex}  
                                                                                                   
\slugcomment{To appear in ApJ}
                                                                                                   
\shorttitle{Collimated particle beams and precession}

\shortauthors{Deshpande \& Radhakrishnan}
                                                    
\begin{document}
                                                                                                   

\title{Further Evidence for Collimated Particle Beams from Pulsars, and Precession}


\author{Avinash A. Deshpande}
\affil{Arecibo Observatory, NAIC, HC-3 Box 53995, Arecibo, PR 00612 and\\
Raman Research Institute, Sadashivanagar, Bangalore 560080, India}
\email{desh@rri.res.in}

\and

\author{V. Radhakrishnan}
\affil{Raman Research Institute, Sadashivanagar, Bangalore 560080, India}
\email{rad@rri.res.in}
                                                                                                   
                                                                                                   

                                                                                                   
\begin{abstract}
    We follow up on our (Radhakrishnan \& Deshpande, 2001: RD01)
    radically different interpretation of the observed structures and
    morphologies in the x-ray observations of the nebulae around young pulsars (PWNe).
    In our general model for PWNe (RD01), originally motivated by the 
    Chandra observations 
    of the Vela X-ray nebula, the bright arcs, the jet-like feature
    and the diffuse components in such nebulae can be explained
    together in detail, wherein the arcs are understood as traces of the 
    particle beams from the two magnetic poles at the shock front.
    We consider this as important evidence for collimated
    particle beams from pulsars' magnetic poles. 
    In this paper, we discuss the variability in the
    features in the Vela X-ray nebula observed by Pavlov et al. (2003), 
    and assess the relevance and implication of our model to the observations 
    on the Crab and other remnants. Our basic picture after incorporating
    the signatures of free precession of the central
    compact object can readily account for the variability and significant
    asymmetries, including the bent jet-like features, in the observed morphologies. 
    The implications of these findings are discussed.
\end{abstract}

                                                                                                   
                                                                                                   

\noindent{\section{INTRODUCTION}}
The superb capabilities of the
Chandra telescope have over the last 6 years revealed several
spectacular images in X-rays
of the nebulae surrounding pulsars, highlighting and resolving the various
spatial structures richly loaded with information about many aspects. 
While most, if not all, of these images bear a remarkable commonality that is
emphasized by their overall symmetric morphology 
about what is readily identified as the direction of the
projected rotation axis, the forms and proportions 
of the components appear to
differ significantly.

An overwhelming majority of the observers and theorists 
interpreting these observations seem to suggest and endorse the following basic picture. 
The jet-like features nearly along the symmetry axis, bisecting the arcs and
the diffuse glow spread about them, are identified with collimated
outflows of relativistic particles along the spin axis of the central compact
remnant, a pulsar. 
The two arc-like features lie along circular rings highlighting
shocks in which the energy of an outflowing equatorial wind is dissipated
to become the source of synchrotron emission for the compact nebula and
the incompleteness of the rings is attributed to preferential Doppler boosting
of the emission in the forward direction. The two rings, if apparent,
 straddle the equator symmetrically, and the deficit of
emission exactly in the equatorial plane is related to the fact that this
is where the direction of a toroidally wrapped magnetic field changes sign
i.e. the field may vanish there. 

    However, Radhakrishnan \& Deshpande (2001, hereafter RD01) have suggested an
    alternative interpretation that differs from this mainstream model
    {\it in practically every aspect}, 
    particularly those regarding the arcs and the jet-like features.  
    The bright arcs, the jet-like feature
    and the diffuse components (for example, in the Vela X-ray nebula) are explained
    together in detail by our ``Rotating Vector model" in which the arcs are understood as
    traces of the particle beams from the two magnetic poles at the shock
    front. We consider this as important evidence for collimated
    particle beams from pulsars, a point that we find necessary to reemphasize.
    
    In this paper, we follow up on the RD01 model
    to address the new clues provided by the variability of various
    features, such as those in the Vela X-ray nebula observed by Pavlov et al. 
    (2001, 2003),
    and assess the relevance and implication of our model to the observations
    on other remnants, such as the Crab. 
    In the next section, we begin with a brief summary of our
    model (RD01) for the x-ray nebulae around pulsars. 
    In section 3, we revisit the Vela nebula story, now incorporating
    signatures of free precession of the central
    compact object, to explain the observed variability.
    We try to model, in section 4, the observed morphology of 
    the X-ray nebula surrounding the Crab pulsar,
    and the significant asymmetries, including the bent jet-like features.
    The implications of these findings are discussed in the last section.

\noindent{\section{THE ROTATING VECTOR MODEL: ORIGIN OF THE ARCS, APPARENT JETS AND THE DIFFUSE GLOW}}

We begin with a brief summary of our general
model (RD01) for the x-ray nebulae around pulsars,  
motivated initially by the Chandra observation of the Vela nebula 
in x-rays showing a well-defined 
pair of bright arcs and the jet-like feature(s) 
surrounded by a diffuse glow of X-ray radiation that is roughly symmetrical over
its bright regions.
Noting the clear separation of the Vela X-ray nebular
emission into two elliptic arcs symmetrically located with respect to the
inferred rotation axis, they are interpreted as the traces of the two
particle-beams from the magnetic poles
on the walls of the cavity $--$ the shocked region $--$ created by the pulsar. 
The visible extent of the arcs arises simply from the spread of the
pitch angles at the shock front. 

In our picture, in stark contrast with prevailing faiths, 
the so-called `jet' is really an apparent one, and 
is a manifestation of a physical flow of particles along 
the {\it magnetic} axis of the star, {\it not} along the rotation axis.
The alignment with the
rotational axis of the jet-like feature bisecting the arcs is simply a
projection effect on the sky plane, as explained before (RD01) and
discussed again below. 

Depending on the spread in their energies, a small fraction of the particles 
from the relativistic particle beams leaving the magnetic poles and proceeding
ballistically outwards to the cavity walls may 
suffer repeated, small latitudinal deviations in their
trajectory through the cavity due to the toroidal field (that they are
carrying out, and) that would be perpendicular to their path, 
and, of course, the rotation axis. In addition to the original
collimated flow ideally amounting to a pencil beam of particles along the rotating
magnetic axis vector, a latitudinally
and progressively elongated flow of particles will now develop on either side of it during
their long passage to the cavity wall. Given that the particles in the flow
components would continue to share the original longitudinal spread, their synchrotron
radiation can be visible to us only during certain rotational longitudes.
We “see” radiation when the projection of the magnetic axis coincides with
that of the rotation axis {\it exactly as in the case of the radio pulse}, but now 
also over an apparently large range of angles in latitude. 
The jet-like appearance would thus be a result of radiation
from particles across the entire latitudinal spread when our sight-line happens 
to be tangential to their motion.
The process that causes some particles to spread latitudinally from their original
path is essentially 
that necessary to deflect them so that
sometimes their radiation is directed towards us.
Wherever the observer, the
apparent jet will appear along the minor axis of the projected ellipses of
the arcs, but the extent over which it is visible will depend on the spread
of particle velocities, as well as the angles that the beams from the two poles make
to the line of sight. 
Since this radiation will occur before the particles reach  
the cavity wall, 
the apparent extent of the jet-like feature is expected to be 
confined to the projected dimension of
the cavity in the latitudinal direction, but could exceed the
extent of even the major axis of the ellipses of the arcs depending on the cavity
shape. When these particles reach the cavity walls,
they will create a diffuse glow around the arc regions but with a greater
spread, {\it exactly as seen in the Chandra image of the Vela nebula}. The spread
of this weak fan beam can be assessed from the size of the diffuse glow, and from the
poor or non-visibility
of the corresponding radiation from the beam of the other
magnetic pole. 

A detailed discussion of our model and its application to the Vela X-ray nebula
can be found in RD01. Here we focus on some of the key features which, we believe, need to be 
re-emphasized. The so-called MHD plasma wind from pulsars, in our picture, 
is essentially in the form
of a pair of collimated relativistic particle beams along the rotating magnetic axis vector.
This is in no way different from that invoked universally to explain radio
pulsar radiation, which
corresponds to only a very tiny fraction of the high energy associated with the particle flow.
Ironically,  several major consequences of such a collimated energetic particle flow, and its manifestations
through unavoidable interaction with immediate and distant surroundings of a pulsar, 
which should have been anticipated, somehow remained
largely unexplored. 
Also, what is not appreciated even after the seminal
paper of Rees \& Gunn (1974) is the {\it inevitability} of the creation
of the cavity by the low-frequency radiation associated with the rotation of
the pulsar at a frequency well below the plasma frequency of the surrounding
medium. Our model merely recognizes and illustrates some of these direct 
consequences/manifestations of the collimated particle beams and the cavity
to explain together the arcs, 
the ``jet" and the diffuse components, as in the case of the Vela X-ray nebula.  
The gratifying agreement of our simulations based on this model (see Fig. 3 of RD01)
with the Chandra observations\footnote{
Image available at\\ http://chandra.harvard.edu/photo/cycle1/vela}, provides 
compelling evidence for
collimated particle beams from pulsars, 
and for the ``apparent" nature of the jet-like feature, in contrast with
any physical jet along the rotation axis of an {\it isolated} pulsar.
The apparent jet-like feature, consistent with
the underlying process, is expected to be 
linearly polarized parallel to itself, and the rotation axis.

The key input parameters for our model are a) the viewing geometry characterized by the inclinations
of the rotation axis with respect to the sight-line and the magnetic axis (i.e. $\zeta$ and $\alpha$,
respectively, where the impact angle $\beta$ is given by $\zeta$ - $\alpha$), b) the pitch-angle distribution
or the spread of radiation occurring at and beyond the cavity wall, c) the latitudinal spread developed
during the passage within the cavity, and d) the shape of the cavity. The first of these can be known
from radio polarization and/or can be estimated from the ellipses associated with the arcs, by fitting
the rotating vector model (as illustrated in Fig. 1 of RD01, based on analytical description as in
Deshpande et al. 1999). It follows that the
ratio of the minor to the major axes will be equal to $\cos \zeta$, and the {\it signed} value of $\beta$
would be apparent from the normalized separation of the nearest arc from the pulsar location.
The pitch-angle distribution can be estimated from the angle between the rotating vector direction
corresponding to the end points of the arc and the sight-line. The third input, the latitudinal spread
may be assessed from the apparent jet extent in units of the projected latitudinal dimension of the cavity.

The shape and the dimension of the cavity created by pulsars are by far the most uncertain parameters. 
This is despite the fact that such a cavity was elaborated in the paper by Rees and Gunn (1974)
for the Crab referred to earlier, and has since formed a part of most, if not all, subsequent
discussions and models of pulsar created nebulae. The arcs do sample
the shape, but only at certain latitudes. The diffuse component does potentially sample a much wider
latitude range, but not fully in most cases. 
The relative intensities of the different components, such as the arc, jet and the diffuse glow, as well
as the width of the pencil beam of particles and the 
position angle (PAo) of the rotation axis projection on the sky-plane
 are the other inputs to the model. A quantitative comparison
with the observations would in principle provide best-fit estimates for most of the mentioned parameters,
except for the shape of the cavity.  
It may not be always possible to assume symmetry in the cavity shape 
about the star's equatorial plane, or even about the rotation axis, since 
the large space velocity of a pulsar can
significantly displace it's location from the point of symmetry, if any, for the cavity. It is not clear how
the cavity shape and size would evolve when a pulsar moves across a significant fraction of the cavity size 
during its life-time. 
Fortunately, in our picture, the arcs provide us important information including about  
the otherwise `unseen' pole. By allowing in the model for unequal values of
$\alpha_1$ \& $\alpha_2$ 
(the half angles of the polar cones associated with the two poles), and similarly for the semi-major axes 
$r_1$ \& $r_2$ of the two elliptical traces, assessment of the apparent asymmetry becomes possible
at least to its first order.

\noindent{\section{REVISIT TO THE X-RAY VELA STORY: PRECESSION INDUCED VARIABILITY ?}}

Now let us take a look at the subsequent observations, 
reported by Pavlov et al. (2001,2003), that show significant 
variability in the locations and the intensities of the 
arcs and the jet-like feature in
the Vela x-ray nebula. The first indication of such variability was 
noted in the two epoch observations by Helfand et al. (2001) 
who found a 5\% brightening of the outer arc within a month, and
suggested a connection with the large glitch (Dodson, McCulloch 
\& Costa 2000) that had occurred a few days prior to their first
observation. 

Initial follow up by Pavlov et al.
(2001), with further Chandra observations separated by 7 months, 
showed changes up to 30\% in the brightness of various features 
of the nebula. Shifts up to a few arcseconds and/or spectral 
changes in the various {\it elements} of the nebula were also 
noticed. The several subsequent Chandra observations, 
providing together a set of 13 epochs 
spread over about two and a half years, also display similar or
stronger variability (as apparent from the impressive animation\footnote{
http://chandra.harvard.edu/photo/2003/\\
vela\_pulsar/animations.html} 
available at the Chandra web-site).
Pavlov et al. (2001, 2003), who reported these
observations, also find a dim, curved, 100" long extension of 
the jet beyond the outer arc, referred by them as an ``outer jet". 
They report that this extension shows particularly strong 
variability, changing its shape and brightness. 
From their analysis of
the image sequence highlighting the so-called ``cosmic fire-hose", 
they ``observed bright blobs in the outer jet
moving away from the pulsar with apparent speeds (0.3-0.6)$c$ and
fading on timescales of days to weeks".
Based on their merged smoothed image, they detect a faint,
strongly bent extension of the outer jet, in addition to
a relatively fainter ``outer counter-jet" that is not apparent
in individual images. Pavlov et al. (2003)
consider the combined action of the wind within the supernova 
remnant, with a velocity of a few x10 km s$^{-1}$, along with
the ram pressure due to the pulsar's proper motion as a 
likely cause for this bend. And they associate the more extreme
bends closer to the pulsar, as well as the apparent side motions
of the outer jet with kink instabilities of a magnetically
confined, pinched jet flow. Their basic picture, however,
is that ``Most likely, these jets are associated with collimated
outflows of relativistic particles {\it along the pulsar's rotation
axes}". It is not surprising then that they liken these features 
to the jets observed in AGNs and Galactic microquasars, 
and hope that studying pulsar jets may shed light on the 
mechanism of jet formation in these as well. 
Our picture of these apparent jets in x-rays 
around isolated pulsars excises the very basis for this hope,
but the observed faint extensions of the jets are not at all
surprising as long as their projected dimension fits within
the projected latitudinal extent of the central cavity.

The most relevant question for us presently is 
what is the underlying 
process responsible for the observed dramatic variability.
Although Pavlov et al. (2003) concentrate on the jet features
in their discussion, we prefer to take a closer look at the 
over-all variability across the nebula (as reported in Pavlov
et al. 2001, and that illustrated by the animation showing
the full set of images). We notice  
significant correlated or systematic variability within the extents
of individual features, as well as across them.
This variability is apparent in the orientation, location, 
brightness of the features.
Regardless of the exact quantitative measure of the 
temporal/spatial correlation between these 
apparently coherent variations, 
it would be far fetched to imagine that the emission received
from sites located several light months apart from each other
co-ordinate their variability, unless they have a common central
origin. We consider two broad classes of central activity 
inducing distinguishable coherent variability of the nebula 
and its fine structures. The first one relates to
any general variability in the collimated particle flows along the
magnetic axis, including those in the particle density, strength of the 
magnetic fields they carry, and the distribution of particle 
energies. Here, we would expect that the resulting variability
in the nebula features to be mainly in form of intensity 
variations, although widths and extents of the narrow features
may also appear to vary. 

In the second class, we consider the
rotational and kinematic history of the pulsar, 
including slow and fast changes 
in its angular velocity, and the effects of its space motion. 
The connection Helfand et al. (2001) suggest between glitch
and the changes in the brightness would be one possible example.
However, the persistent variability in x-ray intensities and locations of most features 
appears to suggest temporally
continuous, and most likely periodic, changes in the angular
velocity of the pulsar, particularly in its direction. 
In this context, we recall the findings
of Deshpande \& McCulloch (1996, hereafter DM96) 
based on their analysis of radio 
data from a long-term dual-frequency monitoring of the Vela pulsar. 

The large fractional variations 
they observed in the pulse intensities at two frequencies 
(635 \& 950 MHz) showed significant mutual correlation, 
and also with the pulse arrival time differences across the
frequencies. The magnitude of such variations and 
the correlations could, in principle, be explained as due to interstellar
refractive scintillations. However, based on their
detection of a significant periodicity of ~330 days (not  
confused with any annual cycle) characterizing these variations, 
and also noting that the sweep-rate
of the polarization position angle measured at 
several epochs shows intriguing and yet unexplained scatter,
DM96 suggested an interesting possibility, namely that
the Vela pulsar is undergoing free precession.

Certain characteristics of the variability in the radio and in
the x-rays appear to be too similar to be considered as 
having different origins.
In view of this and other considerations discussed above,
we interpret the complex variability observed 
in x-rays as a rather direct and natural manifestation of the
free precession of the Vela pulsar. 
If true, the collimated particle beams would also precess,
and the radiation that these particles subsequently produce at 
a given distance from the star will be in accordance 
with the phase of the precession cycle
and with the implied viewing geometric in the past
when they started from the central star. Hence the time taken by
the particles to reach the location of the respective emission,
plus the light-travel time from this location to the observer
will determine the precession phase corresponding to the
emission we sample.
What we will observe is a combined picture of this
differently delayed manifestation of an otherwise
coherent central activity, leading to reduction in the
apparent coherence of variability across the nebula.

To assess this interpretation, we simulate the effect of free-precession
in our model, and the result is shown in the form of an animation
at http://www.rri.res.in/\~desh/VELA2.gif. In Figure 1, we show one sample image
taken from this image time-series. 
In these images, the brightness of the jet feature is artificially
enhanced so as to improve the visibility of the weak jet-like 
feature. 
We have presently assumed the cone
angle and the period associated with the precession to be
5 degrees and 330 days, respectively. We find the remarkable 
qualitative correspondence between the observations and 
our simulations as encouraging. A more detailed quantitative
modeling would need to await access to the observed data
for comparison.
The cavity size in the latitudinal direction is
assumed to be somewhat larger than that in the orthogonal 
direction, in accordance with the extent of the outer
jet feature.  

To summarize, our model based on collimated particle beams
from pulsars, combined with free-precession, can together
explain the detailed morphology, as well as the complex
variability of the Vela x-ray nebula and its components.
We treat this as further strong evidence for collimated
particle beams from pulsars, and for the free precession of
the Vela pulsar.

\noindent{\section{THE X-RAY CRAB STORY: TRACING THE ROTATIONAL HISTORY}}

In our earlier paper (RD01) we had commented that given the similarities observed
between the  morphologies of the surrounding nebulae 
as well as other properties of
the Vela \& Crab pulsars, it would not surprise us if
a similar arc structure is revealed around the Crab pulsar by
observations with improved spatial resolution.
We soon realized after viewing the smoothed Chandra image of 
the x-radiation from the Crab nebula (Weisskopf et al., 2000)
that the arc structure 
is indeed apparent even at the existing resolution, thanks to the
large size of the nebula. 
It is of interest therefore to see if our model
would apply equally well to this case as well.

Using some of the estimates of viewing geometry 
available from the existing radio polarization observation,
along with the RD01 model, we try to simulate the image of
an x-ray nebula surrounding the Crab pulsar. 
The spatial extent of the Crab x-ray nebula is about five times
bigger than that of the Vela pulsar, although comparable in angular
size.
From the Chandra x-ray image, 
we estimate the two radii associated with
the arc structure, and find their ratio to be about 5:2, 
suggesting a very shallow profile for the cavity wall.
A picture in which the pulsar has moved just above the
central neck of an hour-glass shaped cavity seems consistent
with the implied shallowness of the cavity surface, and the
counter-jet extending well outside the extent of diffuse 
emission. 
The visible extent of the arcs in the Chandra image 
implies a wider spread of radiation (more than $\pm 130\deg$)
from the beam trace at the wall, in comparison with that for
the Vela case. Similarly, the latitudinal spread resulting
in the jet-like emission is also wider, judging by
the extent of the feature. The observed jet here is
highly bent, deviating from the symmetry axis (i.e. the
projected rotation axis) on either side of the equator.
These deviations are quite systematic and display striking
antisymmetry with respect to the star's equatorial plane.

Such a systematic pattern on scales of several light-years
between the so-called jet and counter-jet needs attention.
In the commonly endorsed model of the jets
as due to particle flows along the rotation axis,
and moving in opposite direction, the bends in the jets
are believed to be due to exotic instabilities (e.g., Nakamura \& Meier, 2004) developed
due to interaction with the surroundings, combined with the
effect of proper motion of the central star. 
If this picture were to be true, 
the clear antisymmetric signature would
require remarkable continuing communication between the distant parts of 
the particle flows even well after they have parted ways from the
center in opposite directions.

In contrast to this, our model of the apparent jet combined
with possible precession of the star would provide
 a ready explanation
of the observed shape of the ``jet", as well as any variability
in this and other features (Hester et al., 2002; Mori et al., 2004), 
as demonstrated already for the Vela case.  
It is time to recall 
the works of Wills et al. (1982) and Tompkins et al. (1997) in which they examined the
ratio of main-pulse to inter-pulse intensities in gamma-rays
from the Crab pulsar, the former suggesting a 14-year periodicity
associated with the observed variation in the intensity ratio.
Assuming this value of 14-years as the period of precession,
we assess the signature it would imply for the shape of the jet
feature, as well as the overall morphology.     

Figure 2 shows the result of our simulation along the lines
described above, assuming a viewing geometry given by
$\alpha$=86 degrees and $\beta$= -18 degrees, and particle speeds 
to be very close to that of light. 
The precession cone is assumed to be $\pm$12.5 degree wide.
The gratifying qualitative agreement with the Chandra observation\footnote{
http://chandra.harvard.edu/photo/0052/0052\_xray\_lg.jpg} lends
not only strong support to our model of the x-ray nebulae
surrounding young pulsars, but also suggests that the Crab pulsar
too is undergoing free-precession with a period of about 14 years
or so.
 The modelling of the bent jet as due to precession of the central star
 is sensitive to the phase of the precession cycle assumed at the central location,
as well as its variation across the simulated volume. The latter depends linearly
on the distance from the center, and inversely on the
product of the particle speeds with the precession period. The magnitude of the apparent
bend scales directly with the assumed size for the precession cone. 
Although our simulation shown in Figure 2 assumes the precession cone size of $\pm 12.5$ degrees,
a smaller cone size with slower particle speeds would also be consistent with the
magnitude of the bend. Since it is reasonable to expect a spread of energies, we explore
the effect it would have in a simulation shown in Figure 3, where the assumed cone size
is $\pm 7.5$ degrees.

Our Crab nebula simulations indicate that, unlike in the Vela case, the side of the
rotation axis pointing closer to the observer is in the same sense as the star's proper motion
direction.
The extent of the diffuse emission 
depends on the shape and the size of cavity, 
and in the absence of any particular information about these details
a simple ellipsoidal shape is assumed.
Again, several other fine details of interest could be extracted through
 future quantitative comparison of 
our model with the Chandra observations. Further, the implications
of the free precession for the timing data also need to be assessed
carefully. In doing so, the role of glitches in the rotation
history of these pulsars, including their interplay with precession,
would need to be understood, but is beyond the scope of the
present discussion.

It is indeed remarkable the way the rotational history of the 
central pulsar is evident, and can be traced, from these x-ray images of the
surrounding nebula. The application of our model to the two cases 
discussed above allows ready interpretation of the observed
spatial and temporal structure to trace the rotational history
of the central pulsar, and the details of the surrounding cavity.

\noindent{\section{DISCUSSION}}

The cavity shapes suggested by our modeling of
the observed x-ray nebulae might appear to differ significantly 
in the two cases (namely,
the Vela and the Crab) discussed
here. The part of the cavity profile sampled by the particle beams from the
Crab pulsar appears to be significantly shallower than that in the case of 
the Vela pulsar. Given the
age and the proper motion of the Vela pulsar, the relative dimensions of its
x-ray arcs appear to favour an hour-glass shaped cavity, rather than an ellipsoid.
The visibility of the jet-like feature apparently 
extending well beyond the diffuse component is not inconsistent
with both of these components sharing a common origin, i.e. the 
latitudinal spread of a small fraction of the otherwise collimated
particle flow, since the apparent relative brightness of the two components
could differ significantly for intrinsic reasons, as well as
those dictated by the viewing geometry. 

In our picture, the observed extents of the jet-like features, after due
accounting of the projection effect, provide
lower limits for the dimension of the cavity in the latitudinal direction,
which appears to exceed the equatorial dimension, at least in the Vela case.
The apparent asymmetry in the `jet' and the `counter-jet' extents in the Vela
nebula is consistent with our expectation of the associated radiation being
symmetric with respect to the direction of the magnetic axis, modified
by the viewing geometry. If the jet-like
emission were to be due to any physical flow of matter at relativistic speeds
along the rotation axis (e.g. as estimated by Pavlov et al., 2003), the
counter-jet should have been more prominent, both in its extent and its brightness,
given that it would be pointing closer to our sightline.
This is definitely not what is observed, and in fact, this aspect was noted as a cause
for concern by Pavlov at al. (2003). 
Further, such a physical jet flow cannot escape its 
dissipation and consequent termination at the relevant region of the cavity wall 
where a cap-like emission feature should have been apparent, but is not observed.
In their `physical jet' picture, the jet instabilities leading to bending etc.
are expected to be due to and along the proper motion of the star. The Crab case
is clearly inconsistent with this expectation, implying at the least that the 
apparent bending has little to do with the motion of the star. 
These inconsistencies and the other aspects discussed by RD01 argue
strongly against a physical jet along the rotation axis.

Apparent bends in the jet-like features in general, and particularly those
with anti-symmetry about the equatorial plane, are interpreted in our model
as due to possible free-precession. Its successful application to the
Crab and the Vela pulsars, differing in the viewing geometry and rotation
history, is not a chance coincidence. It would not surprise us
therefore, if
the bent jet-like feature in the x-ray nebula
around pulsar B1509-58 (Gaensler et al., 2002) would have a similar interpretation.
The variability of these and other
features is then a natural consequence of such a rotational history,
as illustrated by the multi-epoch observations of the Vela x-ray nebula.
The variability seen in the Crab nebula at optical wavelengths would be another
illustration of the same effect, though on much longer time-scales.
If our assumed precession period of 14 years for the Crab pulsar is
correct, a significant change in the orientation of the jet-like
feature should be expected in the coming years.
As commented earlier, the yet unexplored interplay between the glitches 
and possible precession would determine how the rotation/precession
 history evolves
with time. Hence it is not at all clear whether free-precession, if any,
 would be unaffected through the glitch episodes, and even if it does, whether
any clear signature
of free-precession of the star would be apparent from the radio pulsar
timing residuals obtained after fitting for period glitches and the
associated recoveries commonly observed in young pulsars. 
It remains to be seen if any significant variation in the shape and
intensity of the radio pulses from the Crab pulsar reveals a signature
consistent with the rotational history of the star suggested by 
the above mentioned observations at high energies.

There have been suggestions of free-precession in a few pulsars, but they
are based on apparent changes in pulse shapes and intensities 
(e.g., DM96; Shabanova et al., 2001), or in pulse arrival times (Stairs et al., 2000).
We wish to point out that a direct and independent way of probing 
any changes in the orientation
of the rotation axis of the star, as in precession, is through monitoring
of the polarization position angle (PA) sweep across the radio pulses, and measuring
systematic changes, if any, particularly in the PA sweep-rate (Hari Dass \& Radhakrishnan, 1975). 
We have indeed begun recently
such a monitoring of the Vela pulsar using the Giant Meter-wave Radio Telescope
(GMRT) in India.

To summarize, we find 
 compelling evidence for collimated particle beams from
pulsars, and for free-precession of the Vela and the Crab pulsars, based on
the Chandra observations of the respective x-ray nebulae and their
apparent systematic variability. Quantitative estimates of the intensities
\& the locations of the different elements of the nebulae and their variability,
when compared
in the framework of our simple interpretation, should pave the way to important
clues on the properties of the cavities created by the pulsars, and
the energy spectra associated with the collimated particle beams.

\acknowledgments

\onecolumn
\clearpage

\begin{figure}
\includegraphics[angle=-90,scale=.8]{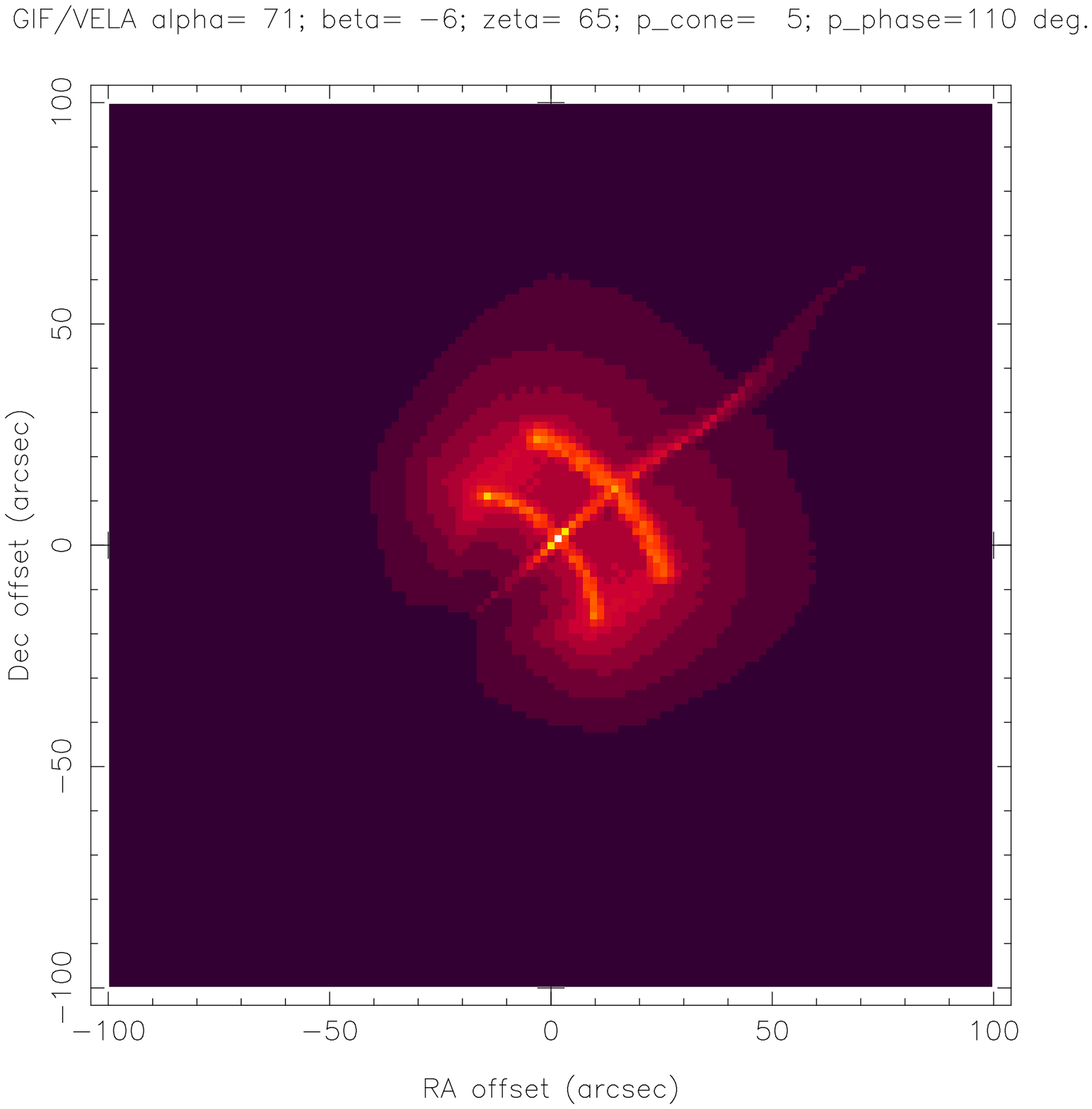}
\caption{A simulated image of the X-ray radiation surrounding the Vela pulsar following the RD01
model, and assuming that the Vela pulsar is undergoing free precession with a period of about 330 days 
and with a precession cone width of $\pm 2.5$ degrees. The intensity of the so-called `jet' component
is enhanced for better visibility. The offsets (0,0) correspond to the pulsar location.
\label{fig:vela_sample}}
\end{figure}

\clearpage

\begin{figure}
\includegraphics[angle=-90,scale=.8]{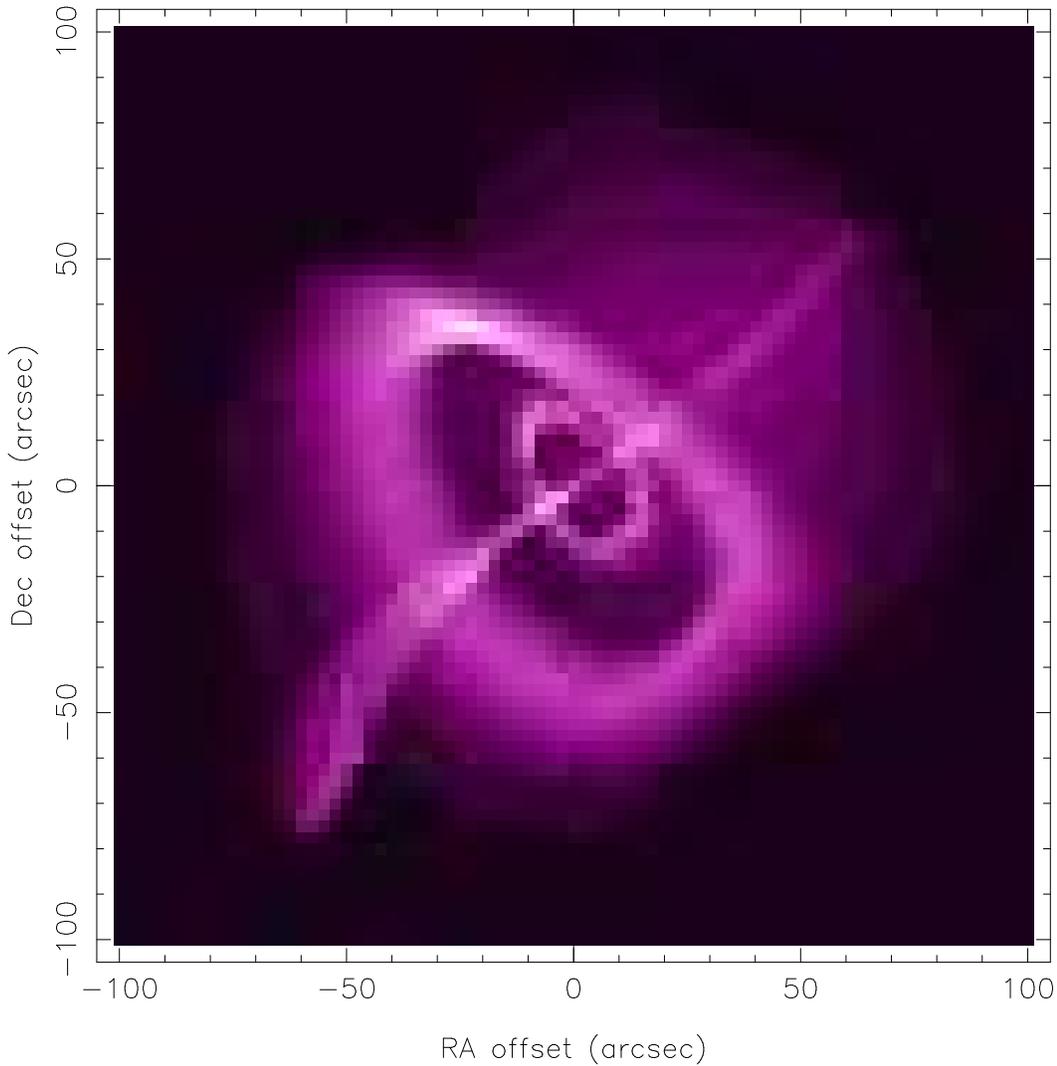}
\caption{Similar to Figure 1, but now showing a simulated image of the X-ray nebula 
surrounding the Crab pulsar.  The assumed precession period for the Crab is 
14 years, and a precession cone width of $\pm 12.5$ degrees. 
The apparent spatial scale of the `jet' bend
is consistent also with other suitable combinations of values for the precession period and
particle speeds. The magnitude of the bend would be consistent with a somewhat 
narrower precession cone size, but with slower particle speeds. 
\label{fig:crab_simul}}
\end{figure}

\clearpage

\begin{figure}
\includegraphics[angle=-90,scale=.8]{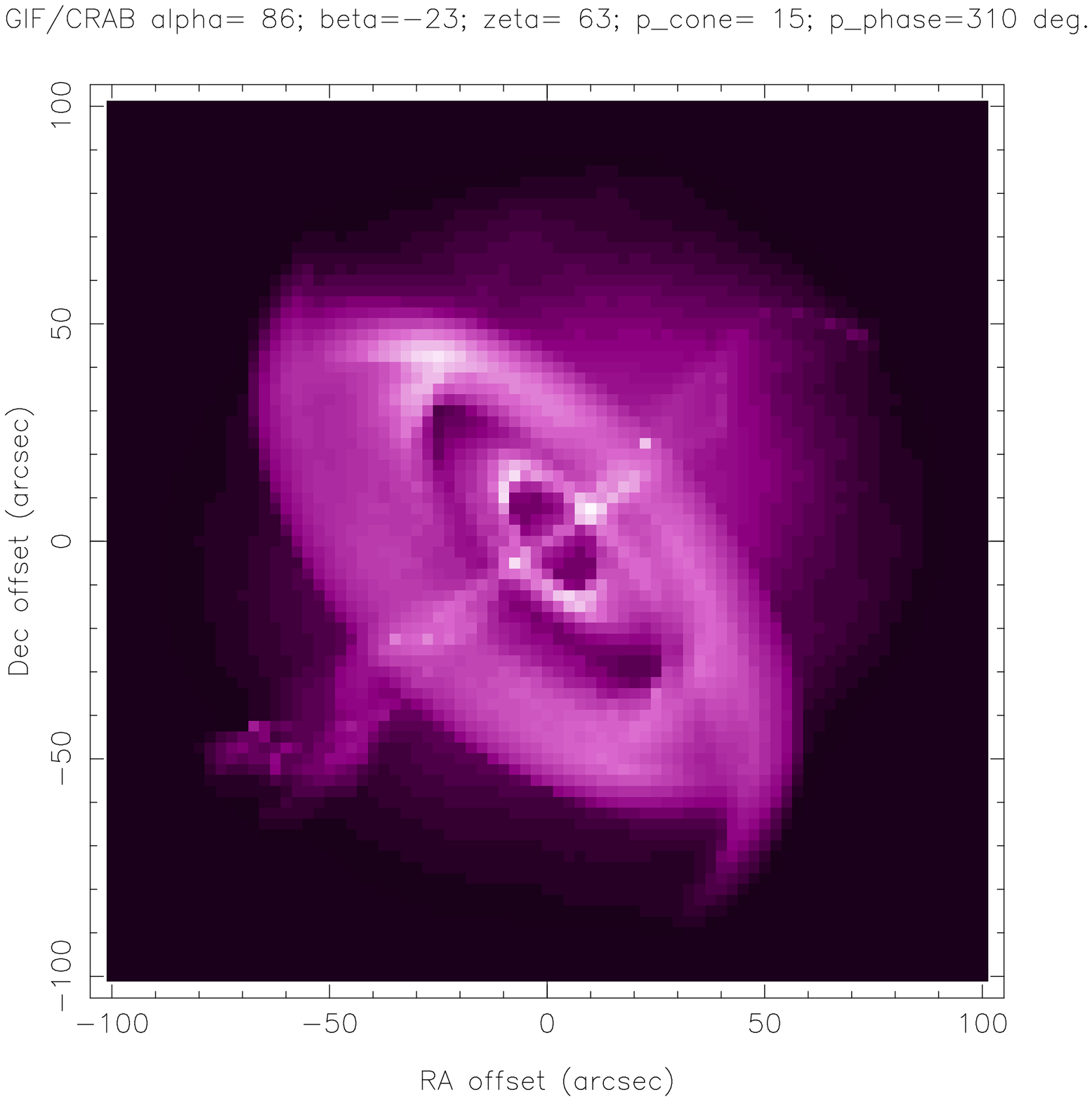}
\caption{Another simulation of the X-ray nebula 
surrounding the Crab pulsar, now illustrating the effect of a spread in particle energies.
A smaller precession cone width of $\pm 7.5$ degrees is used in this case, 
and the assumed combination of the precession phase \& the mean impact 
angle ($\beta$) is suitably adjusted to approximately match the apparent 
viewing geometry with that in Figure 2.
The contribution from lower speed particles to the jet-like feature 
results in relatively sharp apparent bends, comparable with
those seen in Figure 2 for a wider precession cone size combined 
with higher particle speeds. However, the average bend is milder, 
consistent with the reduced values of the cone size. 
The spread in speed (0.1c to c) causes, as expected, a spread in 
the apparent relative orientations/distortions of the arcs traced 
by different speed particle beams, together giving a torus-like 
appearance for the pair of traces, and with the lower speed beams 
showing noticeable anti-symmetric distortion at the tips of
the traced ellipses.
\label{fig:crab_simul2}}
\end{figure}


\begin{thebibliography}{}
\bibitem[]{728} Deshpande, A. A., \& McCulloch, P. M. 1996, ASP Conf. Ser, Vol. 105, Pulsars: Problems \& Progress, eds. S. Johnston, M. A. Walker, M. Bailes (Astron. Soc. Pac., San Francisco), p. 101 (DM96)
\bibitem[]{729} Deshpande, A. A., Ramachandran, R., \& Radhakrishnan, V. 1999, A\&A, 351, 195
\bibitem[]{730} Dodson, R. G., McCulloch, P. M., \& Costa, M. E. 2000, IAU Circ. 7347.
\bibitem[]{731} Gaensler, B. M., Arons, J., Kaspi, V. M., Pivovaroff, M. J., Kawai, N., \& Tamura, K. 2002, ApJ, 569, 878
\bibitem[]{732} Hari Dass, N. D., \& Radhakrishnan, V. 1975, ApJL, 16(4), 135
\bibitem[]{733} Helfand, D. J., Gotthelf, E. V., \& Halpern, J. P. 2001, ApJ, 556, 380
\bibitem[]{734} Hester, J. J., Mori, K., Burrows, D. N., et al. 2002, ApJ, 557, L49
\bibitem[]{735} Mori, K., Burrows, D. N.,  Hester, J. J., Pavlov, G. G., Shibata, S., \& Tsunemi, H. 2004, ApJ, 609, 186
\bibitem[]{736} Nakamura, M., \& Meier, D. L. 2004, ApJ, 617, 123
\bibitem[]{737} Pavlov, G. G., Sanwal, D., Garmire, G. P., Zavlin, V. E., Burwitz, V., \& Dodson, R. G. 2000, BAAS, 32, 733
\bibitem[]{738} Pavlov, G. G., Kargaltsev, O. Y., Sanwal, D., \& Garmire, G. P. 2001, ApJ, 554, L189
\bibitem[]{739} Pavlov, G. G., Teter, M. A., Kargaltsev, O. Y., \& Sanwal, D.  2003, ApJ, 591, 1157
\bibitem[]{740} Radhakrishnan, V., \& Deshpande, A. A. 2001, A\&A, 379, 551 (RD01)
\bibitem[]{741} Rees, M. J., \& Gunn, J. E. 1974, MNRAS, 167, 1
\bibitem[]{742} Shabanova, T. V., Lyne, A. G., \& Urama, J. O. 2001, ApJ, 552, 321
\bibitem[]{743} Stairs, I. H., Lyne, A. G., \& Shemar, S. L. 2000, Nature, 406, 484
\bibitem[]{744} Tompkins, W. F., Jones, B. B., Nolan, P. L., Kanbach, G., Ramanamurthy, P. V., \& Thompson, D. J. 1997, ApJ, 487, 385
\bibitem[]{745} Weisskopf, M. C., et al. 2000, ApJ, 536, L81
\bibitem[]{746} Wills, R. D., et al. 1982, Nature, 296, 723
\end{thebibliography}
\end{document}